\documentstyle[prd,aps,epsf,floats]{revtex}
\draft
\newcommand{\be}{\begin{equation}}
\newcommand{\ee}{\end{equation}}
\newcommand{\ba}{\begin{eqnarray}}
\newcommand{\ea}{\end{eqnarray}}
\newcommand{\nn}{\nonumber}
\newcommand{\ep}{\epsilon}

\newcommand{\plb}{Phys. Lett. B$~$}

\newcommand{\npb}{Nucl. Phys. B}

\def\omegab{\hbox{$\Omega_b$}}
\def\omegabh2{\hbox{$\Omega_b h_{50}^{2}$}}
\newcommand{\lsim}{\mbox{\raisebox{-1.0ex}{$~\stackrel{\textstyle <}
{\textstyle \sim}~$ }}}
\newcommand{\gsim}{\mbox{\raisebox{-1.0ex}{$~\stackrel{\textstyle >}
{\textstyle \sim}~$ }}}
\begin{document}
\twocolumn[\hsize\textwidth\columnwidth\hsize\csname
@twocolumnfalse\endcsname

\thispagestyle{empty}

\title{
\hbox to\hsize{\large Submitted to Phys.~Rev.~D \hfil E-Print
astro-ph/980000}
\vskip1.55cm
Finite temperature effects on cosmological baryon diffusion \\
and inhomogeneous big-bang nucleosynthesis
}

\author{
In-Saeng Suh
\footnote{E-mail: isuh@cygnus.phys.nd.edu,$~~$life@hepth.hanyang.ac.kr}
and G. J. Mathews
\footnote{E-mail: gmathews@bootes.phys.nd.edu}
}

\vspace{0.8 cm}

\address{
\begin{tabular}{c}
Department of Physics,
University of Notre Dame,\\
Notre Dame, IN 46556, USA
\vspace{0.3 cm}
\\
\end{tabular}
}

\date{\today}
\maketitle
\begin{abstract}
We have studied finite temperature corrections to the baryon transport cross 
sections and diffusion coefficients. These corrections are based upon 
our recently computed renormalized electron mass 
and modified state density due to the background thermal bath in the early
universe.
It is found that the optimum nucleosynthesis yields computed using 
our diffusion coefficients shift to longer distance scales by a factor 
of about 3. We also find that the primordial $^4 He$ abundance decreases by
$\Delta Y_p \simeq 0.01$ while $D$ and $^7 Li$ increase.
Effects of these results on constraints from primordial 
nucleosynthesis are discussed.
In particular, we find that a large baryonic contribution to the closure
density ($\omegabh2 \lsim 0.4$) may be allowed in inhomogeneous models
corrected for finite temperature.
 
\vskip 0.5cm
\end{abstract}
\vskip2.2pc]

\narrowtext

\section{Introduction}

It is generally believed that a cosmological phase transition
from quark-gluon plasma to hadronic matter would have occurred 
when the temperature of the universe was about $T \simeq 100 \sim 200$ MeV
\cite{witten,qhpt}.
This quark-hadron phase transition involves a rich variety of physical phenomena, 
particularly if it is first order. 
One possible consequence of this phase transition (or other mechanism \cite{ibbn}
is the formation of isothermal baryon density fluctuations \cite{qhpt,AH,ibbn,AHS}.
In particular, Applegate and Hogan \cite{AH} suggested that, once formed, such
fluctuations could lead to the segregation of neutrons and protons.
Baryon density inhomogeneities
which persist to  $T \sim 0.1$ MeV could affect primordial
big-bang nucleosynthesis (BBN) yields.
The primordial abundances of light elements in such inhomogeneous 
big-bang nucleosynthesis (IBBN) models would be
quite different from those of the standard homogeneous big-bang
nucleosynthesis (HBBN) model. 

In the IBBN model,
neutrons can diffuse relatively easily through the primordial plasma.
Applegate, Hogan and Scherrer (AHS) \cite{AHS} have shown that 
the different diffusion lengths for neutrons and protons could lead to the
formation of high-baryon-density proton-rich regions and low-baryon-density 
neutron-rich regions.
Before the freeze-out from weak reaction equilibrium (at $T \approx 1$ MeV),
there is no segregation between neutrons and protons because the baryons 
only diffuse efficiently during the fraction of the time they spend as 
neutrons. After weak equilibrium freeze-out, however, protons and neutrons
diffuse independently.  Since protons and neutrons have different
diffusion coefficients, the result can be a spatial segregation of these
species. 
 
In view of the importance of using the light-element yields from BBN to
constrain both the baryon-to-photon ratio and various cosmological and    
particle-physics theories \cite{kolb}, such IBBN models must be examined
seriously. It is therefore important to evaluate the diffusion coefficients 
of neutrons and protons as accurately as possible in order to 
quantify the degree to which 
the primordial abundances of the light elements can be modified in the IBBN model. 
In this regard, AHS have derived the baryon diffusion coefficients using a 
mobility formula and the Einstein relation between mobility and the diffusion 
coefficient. After that, Banerjee and Chitre \cite{banerjee} used  
relativistic kinetic theory to calculate the baryon diffusion coefficients 
in the lowest order Chapman-Enskog approximation.

In calculations of the baryon diffusion coefficients, it is necessary to know 
the baryon transport cross sections for scattering with the background primordial 
plasma \cite{AHS}.
However, in all previous works, 
the transport scattering cross sections were calculated in vacuum 
and the electron mass was neglected.
Therefore, in the present work
we take into account finite temperature effects in the calculation
of baryon diffusion coefficients and explore their effects on primordial 
nucleosynthesis.

Finite temperature effects on elementary processes are significant from
the point of view of cosmology and astrophysics. The early universe
is usually described as a hot gas of particles in nearly thermodynamic
equilibrium. Finite temperature effects
enter through the statistical distribution functions which renormalize
the masses and wave functions.
These renormalized masses and wave functions affect scattering processes
and decay rates.

Several authors \cite{donoghue} have generalized
the electron mass and wave-function renormalization to all temperatures
and densities. As an application of finite temperature effects,
Dicus et al. \cite{dicus}, and independently Cambier et al. \cite{cps}, included
the finite temperature effects on weak reaction rates in the calculation of
standard big-bang nucleosynthesis.
They obtained the corrected light-element
abundances and found that the corrections are only of order of a few percent. 
In subsequent work,
Saleem \cite{saleem} included the effects of the electron mass shift at finite 
temperature on BBN, and Baier et al. \cite{baier} examined the finite temperature
radiative corrections to the weak neutron-proton decay rates.
More recently, Fornengo et al. \cite{cwkim} have considered the finite temperature
effects on the neutrino decoupling temperature which is important in the evolution
of the early universe.

In this paper we calculate the baryon diffusion coefficients at finite
temperature and apply these new diffusion coefficients to an IBBN model.
We investigate finite temperature effects on the yields of 
primordial $D,\; ^3 He, \; ^4 He,$ and $^7 Li$ in the IBBN model.
In the calculation of light element yields, 
we use a multi-zone computer code \cite{mathews90} which couples the diffusion 
equation with a nuclear reaction network and the cosmic expansion.

The plan of the paper is as follows. In Section 
\ref{sec:finite}, we discuss how to include finite temperature
effects in the calculation.  We briefly
discuss the effective mass of an electron with special attention to the 
$T \lsim~1$ MeV temperature range which is most relevant for the neutron, 
proton diffusion before and during BBN.
We also evaluate finite temperature effects on the baryon transport cross
section in the scattering of baryons from the background electron-positron plasma.
In Sec.~\ref{sec:diffusion} we calculate the baryon diffusion length 
with new diffusion coefficients for neutrons and protons and compare these to 
those estimated in AHS.
In Sec.~\ref{sec:ibbn} we apply our results to IBBN.
Finally, we summarize our results and discuss the effects of finite 
temperature on light-element constraints from BBN.
We shall employ units in which $\hbar = k_B = c = 1$, except when specific
units must be attached to a result.
  

\section{Finite temperature effects on \\ baryon 
transport cross sections}
\label{sec:finite}

In the early universe where the particles are propagating in 
a thermal bath, their dynamics and 
interactions are modified relative to those in vacuum.
This behavior is systematically 
described in the framework of finite-temperature quantum field theory
\cite{FTQFT}. Finite temperature affects conditions in the early 
universe in the following ways. 
First of all, the free spinors used in the derivation of the cross section
must be replaced by free finite-temperature spinors $u_{T}(p)$ which
describe freely propagating particles in the background thermal bath.
Also, the propagators are modified even at the tree level. Since these break
the Lorentz invariance, they are absorbed into the effective mass 
for the particles. This effect can be evaluated by calculating the self-energy
of the particle in the thermal bath.
At the same time, finite temperature modifies the cross section due to the 
change of propagators and 
spinors in the scattering amplitude and in the distribution functions $f(E)$.

Secondly, the density of final states used in the determination of the cross section
is modified by the background thermal bath as follows:
\begin{equation}
\frac{{\rm d}^3 p\,'}{(2 \pi)^3 2 E'}
\longrightarrow
\frac{{\rm d}^3 p\,'}{(2 \pi)^3 2 E'} [1 - f_F(E')] ,
\end{equation}
where $f_F(E')$ is the Fermi-Dirac distribution at temperature $T$ for final energy $E'$.
This takes into account the states already occupied by fermions in the thermal bath.
However, note that since we are still interested in single particle scattering,
we do not make a thermal average over initial states. 
 
The dynamics of electrons in a thermal bath is
modified by the electromagnetic interactions with background photons
and electrons themselves. Therefore, the effect of the thermal bath
on the propagation of an electron is expressed by calculating
the electron self--energy in the presence of the ambient $e^+$, $e^-$
and $\gamma$'s \cite{donoghue}.
The temperature corrected electron physical mass is then
obtained by evaluating the renormalized propagator and then finding the zero of its
inverse. Thereby, the temperature dependent physical mass of an electron 
is given as \cite{donoghue}
\ba
m_{T}^2 &\equiv& m^2 = E^2-\vec{p}\,^2 \nn \\
&=& m_0^2 + \frac{2}{3}\alpha \pi \, T^2 +  \frac{4}{\pi} \alpha m_{0}^2 {\cal B}(x) + 
\frac{\alpha}{2 \pi^2} m_{0}^2 {\cal J}_A (p) ,
\ea
where $m_0 = 0.511$ MeV is the electron rest mass in vacuum and
$\alpha=e^2 /\hbar c$ is the electron fine structure constant.
The function ${\cal B}(x)$ with $x = T/m_0$, is defined as \cite{donoghue}
\be
{\cal B}(x) \equiv \int_{1}^{\infty} ds \, \frac{\sqrt{s^2 - 1}}{e^{s/x} + 1} .
\ee
In Ref. \cite{cwkim} it was shown that the function ${\cal J}_A (p)$ in the 
fourth term in Eq. (2) is negligible for $T \sim ~$ MeV. 
We therefore neglect this term in our analysis.

Eq. (2) is valid for all temperature \cite{saleem}. 
It gives the correct result $m_{T} = m_0$ at $T=0$.
Around $T \sim m_0$, however,
the third term becomes important and has to be taken into
account [for example, ${\cal B}(x=1) \simeq 0.543$].
It also has been shown in \cite{cwkim,isuh} that the thermal corrections to the
electron mass at $T \sim ~$ MeV are sizeable. At $T=1$ MeV
the electron mass increases by 4.1\%; and at $T=2$ MeV the correction
is as large as 16\% \cite{cwkim}.

Now we investigate the transport cross section in electron--hadron scattering 
at finite temperature.
Rosenbluth \cite{rosenbluth} first calculated the electron-hadron differential
cross section under the assumptions that the
electron is ultrarelativistic($m \ll E$)  
in the rest frame of  the incoming hadron. This treatment takes into account the
internal structure and anomalous magnetic moment. 
Applegate, Hogan and Scherrer \cite{AHS} used the Rosenbluth formula in the 
calculation of the neutron diffusion coefficient, $D_{ne}$, 
with the assumption that the electron energy 
is much less than neutron mass $M$. They obtained a constant transport
cross section $\sigma_{t}$ for the vacuum interaction between an electron and 
a neutron \cite{AHS},
\be
\sigma_{t}^{AHS} = 3 \pi \left(\frac{\alpha \kappa_n}{M}\right)^2 
\simeq 8 \times 10^{-31} \;cm^2 ,
\ee
where $\kappa_n = -1.913$ is the anomalous magnetic moment of the neutron.
However, at $T \sim 1~$ MeV,
the dynamical properties of electrons and photons in this thermal bath would be 
changed. (But neutrons would not be affected by the background thermal bath since
their mass is  nearly 2000 times the mass of the electron.) 
We therefore have to take into account the effect of finite temperature on the
interaction
between electrons and hadrons.   

Recently, we calculated the temperature-dependent baryon transport scattering 
cross sections at finite temperature in the early universe \cite{isuh}.
The transport cross section $\sigma_t (T)$ of the scattering process is defined by
\be
\sigma_t (T) = \int d \bar{\sigma} (1 - cos\theta'),
\ee
where $\theta'$ is the scattering angle and 
$d \bar{\sigma}$ is the differential cross section including the thermal phase space. 
In the rest frame of the incoming neutron, 
$k^{\mu} = (k^{0}, k^{i})$, where $k^{0} = \ep = M$, and $k^{i} = 0$, 
integrating Eq. (5) over scattering angle, we obtained the temperature-dependent
transport cross for electron-neutron scattering
\be
\sigma_{ne} (T) = \frac{1}{2 \pi} \frac{m^2 M}{|\vec{p}|^{2}} \int_{E'_{min}}^{E'_{max}} 
d E' (1 - \beta(E')) \bar{|{\cal M}|^2} {\cal S}(E',\ep') ,
\ee
where $m$ is the mass of electron, $M$ is the mass of neutron, 
$|\vec{p}|^{2} = E^2 -m^2$, and
\be
\beta(E') = \frac{E'(M+E) - m^2 - M E}{|\vec{p}||\vec{p}'|} . 
\ee
The statistical factor ${\cal S}(E',\ep')$ is
\be
{\cal S}(E',\ep') = [1 - f(E')][1 - f(\ep')] ,
\ee
where $f(E)$ is the Fermi-Dirac distribution function.
For a given initial electron energy $E$, the kinematical limits on the final state
energy of the electron, $E'_{min}$ and $E'_{max}$, can be determined from the constraint
$|cos \theta'| \leq 1$. The squared spin-averaged scattering matrix element in Eq.
(6) is
\be
\bar{|{\cal M}|^2} = 2 \pi^2 \left(\frac{\alpha \kappa_{n}}{m M}\right)^2 
\left(\frac{1}{2} - \frac{E E' + m^2 }{M (E'-E)}\right) .
\ee

For electron-proton scattering, the most important scattering mechanism
is Coulomb scattering and the differential cross section is given by the Mott formula. 
With the assumption that the electron energy is much less than the proton mass, the
Coulomb transport cross section is \cite{qed},
\be
\sigma_{pe} (T) = 4 \pi \alpha^{2} \left( \frac{E}{E^2 - m^2} \right)^2 \Lambda(T)
[1 - f(E)].
\ee
Because Eq. (10) diverges at small angles,
the usual approximation is to truncate the angular integration at an angle given
by the ratio of the Debye shielding length $\lambda_D = (T/e^2 n_e)^{1/2}$
to the thermal wave length $\lambda_{th} = (2 \pi  / m T)^{1/2}$ \cite{banerjee,kinetic}.
This defines to  Coulomb logarithm $\Lambda (T) = ln (\lambda_D / \lambda_{th})$.
With this we can evaluate numerically the transport cross sections for a given initial
electron energy.

In the early universe, the number density and
energy density of electrons is given by \cite{kolb}
\ba
n_e (T) &=& g_e \int \frac{d^3 p}{(2 \pi)^3} f(E), \\
\rho_e (T) &=& g_e \int \frac{d^3 p}{(2 \pi)^3} E f(E) ,
\ea
where $g_e$ is the electron degeneracy.
For an electron in thermal equilibrium, the phase space occupancy $f(E)$ is given
by
the Fermi-Dirac distribution
\be
f(E) = \frac{1}{e^{(E-\mu)/T} + 1} ,
\ee
where $\mu$ is the electron chemical potential. 
But since the chemical potential $\mu$ is small in the early universe,
$\mu/T \lsim 10^{-9}$ \cite{dicus}.
we can ignore it in the numerical computations
of the average electron energy. 
For the case in which the temperature dependent electron mass can not be neglected,
we can obtain the average electron energy $\langle E \rangle = \rho_e / n_e$ 
from Eqs. (11) and (12).
Since this initial electron energy $\langle E \rangle$ can be a good
approximation in both relativistic and nonrelativistic limit \cite{isuh}, 
we will use it as the initial electron energy
in numerical calculations of the scattering cross section.
Fig. 1 shows the baryon transport cross section as a function of $x = T/m_0$
for the initial electron energy $E = \langle E \rangle$. 
For $T \lsim 1$ MeV, the electron-neutron transport cross section is not constant
but increases as temperature decreases \cite{isuh}. 

The change of mass of the particle could modify its
contribution to the energy density of the universe and therefore
the expansion rate, but in this paper we will not consider this effect.
In addition, the modification of the electron mass
also changes the relationship between the neutrino and
photon temperatures.
The neutrino temperature differs from the photon temperature because 
neutrinos decouple at a temperature $T_d \simeq 1$ MeV. 
Therefore, they do not share in the entropy release 
from electron-positron annihilations which occurs at $T \lsim 0.5$ MeV. 
The modified relationship between the neutrino and
photon temperature is given by \cite{dicus}
\be
T_{\nu} = T_{\gamma} \left[\frac{4}{11} \, \zeta (w) \right]^{1/3}
\ee
where the function $\zeta(w)$ with $w = m / T_{\gamma}$ is defined by 
the integral
\be
\zeta(w) = 1 + \frac{45}{2 \pi^4} \frac{1}{w^4} 
\int_{1}^{\infty} ds \frac{s \sqrt{s^2 - 1}[s + (s^2 - 1)/3 s]}{e^{s/w} +1}
\ee  
In Eq. (14), we have neglected the density and pressure corrections \cite{heckler}
which give a small correction to the primordial $^4 He$ abundance.

\section{Cosmological Baryon Diffusion}
\label{sec:diffusion}

Baryon density inhomogeneities  affect primordial nucleosynthesis
through neutron-proton segregation. This segregation is determined by the 
different neutron and proton diffusion lengths 
during the nucleosynthesis epoch \cite{qhpt,AH,ibbn,AHS}.
At high temperatures the diffusion lengths of neutrons and protons are 
equal because these particles intertransmute rapidly through weak interactions.
After weak decoupling (at $T \approx 1$ MeV) neutrons and protons are no longer 
in equilibrium with respect to weak interactions.
As a result, they retain their identity so that
diffusive segregation can take place.
Neutrons are scattered by electrons and positrons (through the interaction of
their magnetic moments) and by protons due to nuclear interactions.
Protons, on the other hand, undergo Coulomb scattering with electrons and 
are also scattered by neutrons. Other minor scattering mechanisms,
such as neutron-photon scattering, neutron-neutron scattering, etc., 
may be neglected \cite{AHS}. 

AHS \cite{AHS} calculated the diffusion coefficients using a mobility formula 
together with the Einstein
relation between mobility and the diffusion coefficient.
They used a constant electron-neutron scattering cross section [Eq. (4)]
and included a relativistic Maxwellian distribution for background light particles.
Banerjee and Chitre \cite{banerjee} calculated the diffusion coefficients 
in the lowest order Chapman-Enskog approximation within the framework 
of relativistic kinetic theory. 
They showed the equivalence between the expression for the diffusion
coefficients given by the mobility formula and that derived from relativistic
kinetic theory for a dilute neutron gas diffusing through electrons at low temperature.
Kurki-Suonio, et al. \cite{kurki} applied the new diffusion coefficients to IBBN and
showed that the abundance curve shifted towards slightly larger distance scales. 
In the same spirit it is also of interest to investigate the effects of finite
temperature on the baryon diffusion coefficients. 

In relativistic kinetic theory \cite{kinetic}, the diffusion coefficient of particle
$i = n, p$
(neutron, proton) moving through a gas of light particles $j = e$ (electron), 
is given by \cite{banerjee,kinetic,kurki} 
\be
D_{ij} = \frac{3}{8} \sqrt{\frac{\pi}{2}} \frac{1}{n_j \sigma_{ij} (T)}
\frac{z^{1/2} K_{2} (z)}{K_{5/2} (z)} (1-x_i) ,
\ee
where $z = T/m$ and $x_i$ is the fraction of particles $i$  
and can be neglected \cite{kurki}. 
$\sigma_{ij} (T)$ is the transport cross section for the scattering of particles $i$ by
particle $j$ at temperature $T$. The
modified Bessel functions, $K_2$ and $K_{5/2}$, 
are given by 
\be 
K_{2}(z) = \frac{1}{z} \int_{1}^{\infty} dk \, k \sqrt{k^2 - 1} \, e^{-k/z},
\ee
and
\be
K_{5/2}(z) = \sqrt{\frac{\pi}{2}} \frac{\sqrt{z}(1 + 3 z + 3 z^2)}{e^{1/z}} .
\ee
In Eq.~(16), $n_j(j=e)$ is the total density of electrons and positrons.
At high temperature, $m_0 \lsim T$, the electron density is given by the 
Fermi-Dirac distribution from Eq. (11).
However, when the temperature is low enough that the net electrons dominate over 
the thermal electron-positron pairs, nonrelativistic Maxwell-Boltzmann statistics 
can be used to obtain the electron density. 
Then, overall charge neutrality requires that the net electron density be equal
to the total density of protons.
With this requirement, we can replace the electron chemical potential
with the baryon-to-photon ratio $\eta$ in a nucleosynthesis calculation. Thus,
we obtain 
\cite{kurki}
\be
n_e = \left[2 \left(\frac{m}{\pi} T \right)^3 e^{-2m/T} +
 (\eta X_p n_{\gamma})^{2} \right]^{1/2} ,
\ee
where $X_p$ is proton mass fraction and
$n_{\gamma}$ is the photon number density.
Note that the relative difference between the nonrelativistic Maxwell-Boltzmann and
Fermi-Dirac statistics is small in the temperature range of interest.
Therefore, we can use Eq.~(19) in the calculation of baryon diffusion coefficients.
 
The neutron-proton diffusion coefficient is given by \cite{AHS,banerjee}
\be
D_{np} = \frac{3 \sqrt{\pi}}{4} \sqrt{\frac{T}{M_p}} \frac{1}{n_p \sigma_{np}}
(1 - x_n),
\ee
where $M_p$ is the proton mass, $n_p$ is the proton number density, and
$\sigma_{np}$ is the neutron-proton scattering cross section.
For our numerical calculations, we take the scattering 
cross section for s-wave neutron-proton scattering given in Ref. \cite{AHS}.
As is Kurki-Suonio et al. \cite{kurki},
we use Eq. (20) in the calculation of baryon diffusion coefficients
even though it comes from the Chapman-Enskog diffusion coefficient 
for classical hard-sphere scattering.

Finally, the effective neutron diffusion coefficient $D_n$ is 
given by
\be
D_{n}^{-1} = D_{ne}^{-1} + D_{np}^{-1},
\ee
and the effective proton diffusion coefficient $D_p$ is
\be
D_{p}^{-1} = D_{pe}^{-1} + D_{pn}^{-1} ,
\ee
where $D_{pn}$ is the coefficient for the scattering of protons from neutrons.
This quantity is related to $D_{np}$ by \cite{mathews90}
\be
D_{pn} = \left( \frac{X_p}{X_n} \right) D_{np} ,
\ee
where $X_n$ and $X_p$ are the local neutron and proton mass fractions, respectively.
(Note that this corrects the typographical error in Eq.~(5) of Ref. \cite{mathews90}
in which $X_n$ and $X_p$ are reversed.)
Eq.~(23) corrects $D_{np}$ for the fact that the roles of the neutrons and 
protons are interchanged in the scattering leading to proton diffusion.

From Eqs.~(21) and (22), we can compute the comoving diffusion length of baryons 
in the early universe.
Here we take the comoving diffusion length relation from Refs. \cite{AHS,applegate}
in order to compare with those of AHS.
Fig. 2 shows the comoving neutron and proton diffusion lengths as a function of
temperature, where we have normalized the scale factor $R(t)$ to $R = 1$ when the 
neutrino temperature $T_{\nu}$ is $1$ MeV.
Our results are compared with the results of AHS for the case of 
$\eta_8 = 0.1 \; (\eta_8 = 10^8 \eta \simeq 0.67 \, \omegabh2)$,
where $h_{50}$ is the Hubble constant in units of $H_0 = 50 \, km \, s^{-1} \,
Mpc^{-1}$.

The comoving baryon diffusion length at finite temperature is shifted to longer length
scales by about a factor of 3 compared to the AHS estimate except 
the neutron diffusion length near $T \simeq 0.1$ MeV.
The reason that both AHS and our comoving neutron diffusion lengths are
equal around $0.1$ MeV is due to the fact that $D_{ne}^{FT}$ 
(where the superscript $FT$ means that it was calculated with finite temperature 
corrections) intersects $D_{ne}^{AHS}$ around $0.1$ MeV. 
From these results, we see that finite temperature effect
on baryon diffusion gives results which are roughly equivalent to those obtained 
using a smaller value of $\eta$.

\section{Applications to Inhomogeneous Big-Bang nucleosynthesis} 
\label{sec:ibbn}

As primordial abundance measurements become more refined, it may indeed turn out
that the standard BBN model is no longer capable of fitting all of the 
observed abundances \cite{mathews98}. 
Indeed, if one adopts the currently favored lower $D/H$ abundance 
of \cite{tytler} together with the low $Y_p$ of \cite{olive}, a discrepancy 
may always exist \cite{hata}.
In this context, it is worthwhile to consider alternative
cosmological nucleosynthesis models. One of the most extensively investigated
possibilities is that of an inhomogeneous baryon density distribution
at the epoch of nucleosynthesis.  
Such studies were initially motivated by speculation \cite{witten,qhpt,AH}
that a first order quark-hadron phase transition (at $T \simeq 100 \sim 200$ MeV) 
could produce baryon density inhomogeneities as baryon number was
trapped within bubbles of shrinking quark-gluon plasma. 
Other mechanism also exist \cite{ibbn} for the generation of baryon inhomogeneities.

The abundances of primordial nucleosynthesis could be affected by 
these QCD motivated baryon inhomogeneities.
There have been a number of studies on this subject \cite{qhpt,afm}.
Most studies in which the coupling between the baryon diffusion and
nucleosynthesis has been properly accounted for \cite{ibbn,afm}
have concluded that the upper limit on
$\omegabh2$ is virtually unchanged when compared to the upper
limit on $\omegabh2$ derived from standard HBBN,
where the allowed range for $\omegabh2$ in HBBN is
$0.04 \lsim $\omegabh2$ \lsim 0.08$ \cite{orito,mathews96,copi}.
However, in \cite{orito} and \cite{mathews96} it was shown that there exists 
a region of the parameter space for inhomogeneous models in which a somewhat 
higher baryonic contribution to the closure density is possible than that 
allowed in standard HBBN models. 
They found that a baryonic contribution as high as
$\omegabh2 \leq 0.13 (0.2)$ (nearly twice the HBBN upper limit) is possible for the
spherical(and cylindrical)-shell fluctuation geometry.

The purpose of this paper is, therefore, to estimate finite temperature effects
on the IBBN model with our new diffusion coefficients and to investigate constraints on
$\omegabh2$.
The calculations are based upon the coupled diffusion and
nucleosynthesis code of Mathews et al. \cite{mathews90} 
with an initially square-wave spherical inhomogeneity.
Since our study is focused only on finite temperature effects 
(and to compare directly with results of \cite{mathews90,kurki}), 
we do not include other effects \cite{orito} such as ion diffusion, Compton drag,
other fluctuation shape geometry, and finite temperature corrections in weak interactions
\cite{dicus} which will also change BBN yields.

In order to describe the IBBN a number of parameters are introduced.
The initial fluctuation shape is taken as regions of different 
baryon density separated by a sharp boundary. The physical quantities
which characterize the fluctuation are the ratio of baryon number density,
$R$, the mean separation between nucleation sites, $r$, the relative volume fraction
$f_V$ occupied by the high-density zones, and the total baryonic contribution to 
the closure density $\omegab$. This later quantity we relate to the baryon-to-photon
ratio $\eta$ by assuming a present background temperature of $2.75$ K and a Hubble
constant $h_{50}$. 

We have performed the calculations with a range of parameters, and compare
those results with the primordial abundances.
We choose $R=10^6$ and $f_{V}^{1/3} = 0.25$ for the high density region.
For all calculations, we use three neutrino species and
the neutron half-life $\tau_n = 10.70$ minutes \cite{mathews90}.
The variable parameters in the calculations are the average separation distance 
between fluctuations $r$ and the total average baryon-to-photon ratio
$\eta$.

Fig. 3 shows the thermal evolution of the neutron and proton diffusion coefficients
during the nucleosynthesis epoch for the parameters, $R = 10^6, \; f_{V}^{1/3} = 0.25$
and for the case of $\omegabh2 = 0.1$ for $r=100$ m.
At high temperature ($ T \gsim 0.1 \, MeV$), both $D_{n}^{FT}$ and $D_{p}^{FT}$  
are much larger than those of AHS. 

We also note that in our calculations neutron diffusion coefficients and 
proton diffusion coefficients are nearly 
independent of $\omegabh2$ because the diffusion coefficients are dominated by 
electron scattering. 
The spike in the proton diffusion coefficient of AHS indicates neutron back diffusion
\cite{MF}.
The absence of a spike in our calculations indicates that there is little 
neutron back diffusion when finite temperature effects are included. 
This is because of the smaller $D_n$ at low temperature.

The resulting abundances as a function of the distance scale of the inhomogeneity
are shown in Figs. 4 and 5. 
The results are somewhat different from those of previous works 
\cite{mathews90,kurki}.
It is found that the optimum nucleosynthesis yields computed using   
our diffusion coefficients is shifted to longer distance scales by a factor
of about $3$. We also can see that the value of the $^4 He$ abundance 
significantly decreases by $\Delta Y_p \simeq 0.01$. There is also an increase in 
the primordial $D$ and $^7 Li$ abundance.
 
In order to see the allowed region for values of $\omegabh2$ and $r$,
we adopt the following observational constraints on the $^4 He$ mass fraction
$Y_p$ from Ref. \cite{orito}:
\be
0.226 \leq Y_p \leq 0.247 .
\ee
For the primordial deuterium abundance, we adopt the constraints
including implications of the possible detection of a high deuterium abundance
in Lyman-$\alpha$ absorption systems \cite{songaila} as an upper value,
$D/H = 1.9 \pm 0.4 \times 10^{-4}$.
The average lower value of $D/H$ \cite{tytler} is given
as $D/H = 2.4 \pm 0.9 \times 10^{-5}$.
Finally, we take a conservative upper limit on the primordial lithium abundance
of $^{7}Li/H \leq 1.5 \times 10^{-9}$ \cite{orito} as well as
the upper limit to the lithium abundance suggested in the detection
(along with the possible depletion) of lithium in stellar atmospheres,
$^{7}Li/H \leq  3.5 \times 10^{-10}$ \cite{copi}.

Fig. 6 illustrates the allowed parameter region in the $\omegabh2$ versus
$r$ plane for condensed sphere fluctuations for the adopted light-element
abundance constraints. We note that, for the preferred low $D/H$ abundance, 
no agreement is possible in the HBBN model without expanding both 
$Y_p$ and $Li$ limits. However, in the IBBN model a concordance region is
possible between $D/H$ and $Y_p$ with the larger $Li/H$ abundance upper limit.

We can see that the primordial lithium abundance
crucially constrains the allowed values of $\omegabh2$.
This is because the back diffusion of neutrons which can destroy lithium \cite{MF}
is hindered by the new diffusion coefficients.
Nevertheless, by adopting a conservative upper limit to $Li/H$, large values 
of $\omegabh2$ are allowed.
The optimum fluctuation distance scales
($20 \sim 60 \, m$) are about a factor of 3 larger than those
implied by calculations using the AHS diffusion coefficients. 
With the larger value for the $D$ constraint \cite{songaila},
allowed values of $\omegabh2$ are nearly unchanged when compared to previous works
\cite{mathews90,orito,mathews96} 
However, for the smaller value of the $D$ constraint \cite{tytler}
(together with a high lithium constraint),
we obtain allowed values of $\omegabh2$ as large as $\lsim~0.4$ !


\section{Conclusions}
\label{sec:conclusion}

We have studied the leading finite-temperature effects on baryon diffusion
and nucleosynthesis in the early universe. 
The major motivation has been 
to quantify the degree to which finite-temperature effects could 
affect the nucleosynthesis constraints.
We have calculated the baryon diffusion lengths based upon our previously 
calculated scattering cross sections \cite{isuh}
which included:
(1) finite temperature Dirac spinors
which are recast into the form of an effective electron mass;
(2) finite temperature modifications to the phase space
distribution of the electrons; and 
(3) the modified neutrino temperature.
 
The baryon diffusion coefficients which affect baryon
inhomogeneities before or during big-bang nucleosynthesis are changed
significantly by our temperature dependent electron-hadron transport cross sections.
We have reevaluated the upper limits to $\omegabh2$ in inhomogeneous primordial
nucleosynthesis models which take into account finite temperature effects.
The optimum conditions are shifted to larger fluctuation distance scales 
by a factor of about 3 in condensed spherical fluctuation geometry.
It also is found that the value of $^4 He$ decreases by
$\Delta Y_p \simeq 0.01$.
The limits on the baryon to photon ratio are constrained, however, by increased
$^{7}Li$ which limits the allowed range of $\omegabh2$.

With the larger value of the $D$ constraints, 
allowed values of $\omegabh2$ are nearly unchanged when compared to previous works.
However, for the preferred smaller value of the $D$ constraints,
it is possible to have an allowed region of $\omegabh2$ as large as 0.12 to 0.4
in inhomogeneous models.
 
\vspace{1cm}

{\bf Acknowledgments.}
\noindent
The author would like to thank Prof. M. Orito for his careful reading of an early
version of this manuscript.
I.S.S. also acknowledges the Korea Research Foundation (KRF) for financial support.
This work supported in part by DOE Nuclear Theory Grant DE-FG02-95ER40934.
 


\newpage
\begin{center}
\begin{large}
FIGURE CAPTIONS
\end{large}
\vspace{5mm}\
\end{center}
 
\begin{itemize}
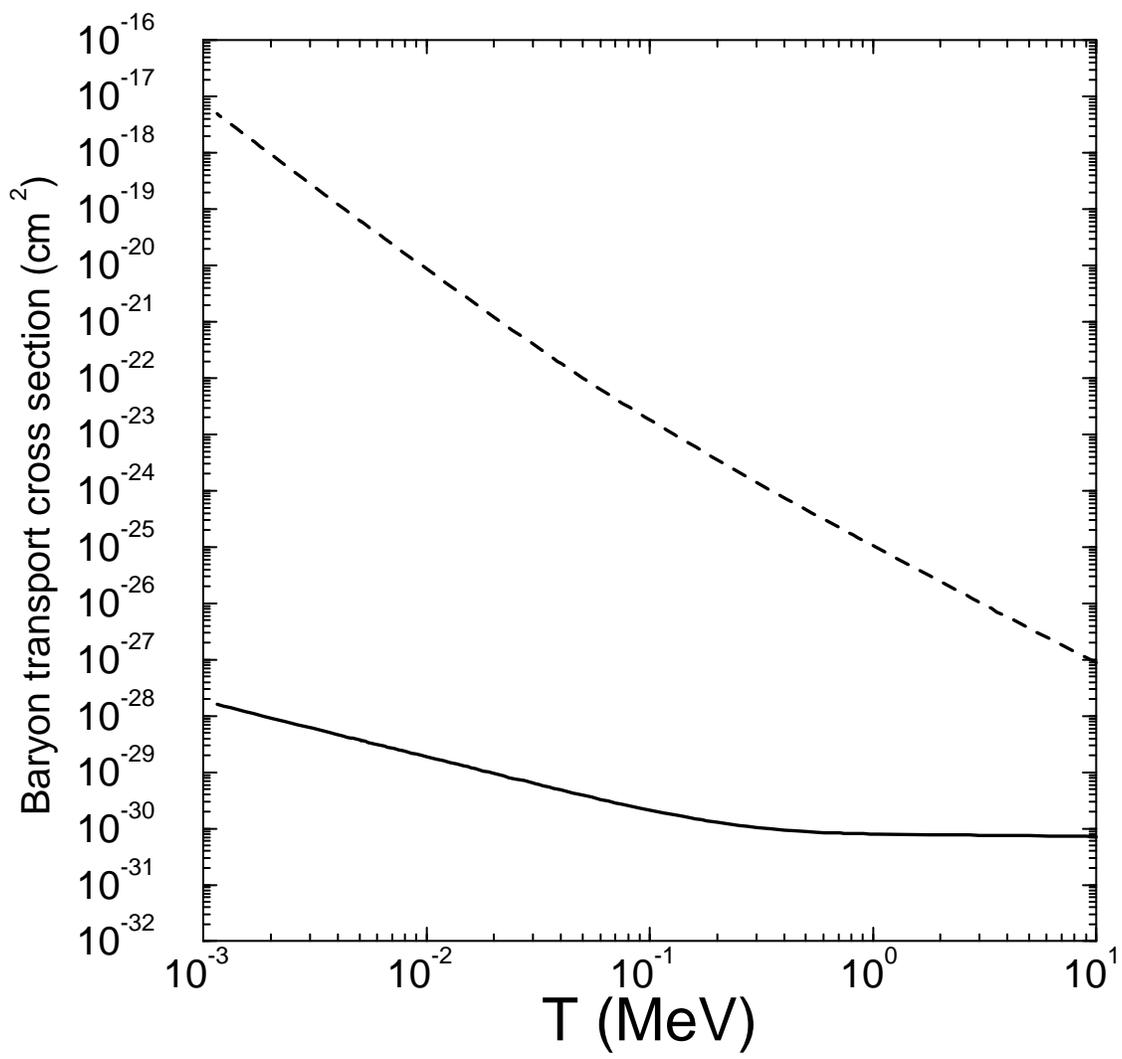
\item [{\rm Fig. 1}]
The baryon transport cross section as a function of $x = T/m_0$
for the initial electron energy $E = \langle E \rangle$.
The solid line shows the temperature-dependent electron-neutron transport cross section.
The dashed line denotes the electron-proton transport cross section.

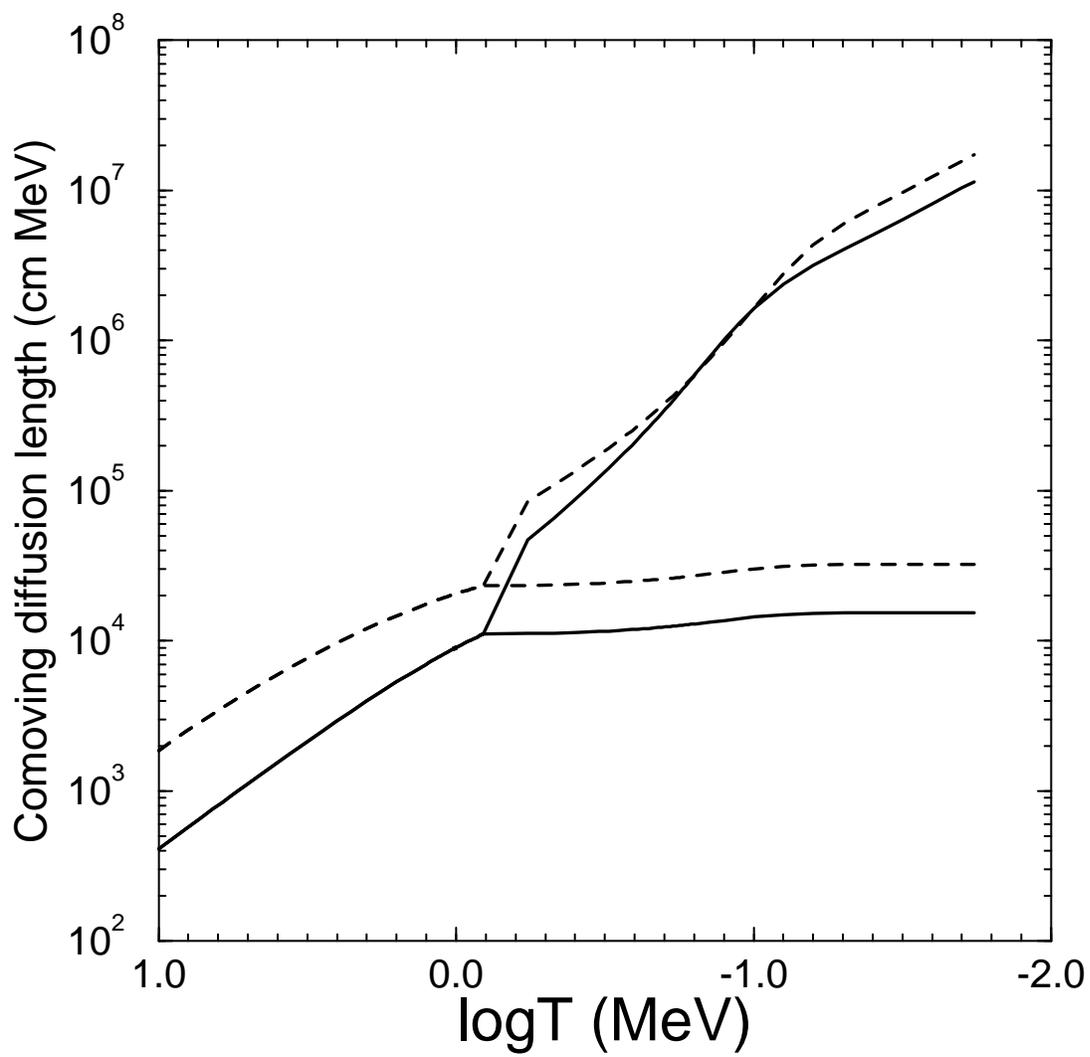
\item [{\rm Fig. 2}]
Comoving neutron and proton diffusion lengths as a function of
temperature. Here we normalize the scale factor $R(t)$ to $R = 1$ when the
neutrino temperature $T_{\nu}$ is $1$ MeV.
The solid line corresponds to the result of AHS. The dashed line is the
comoving baryon diffusion length which takes into account finite temperature effects.
 
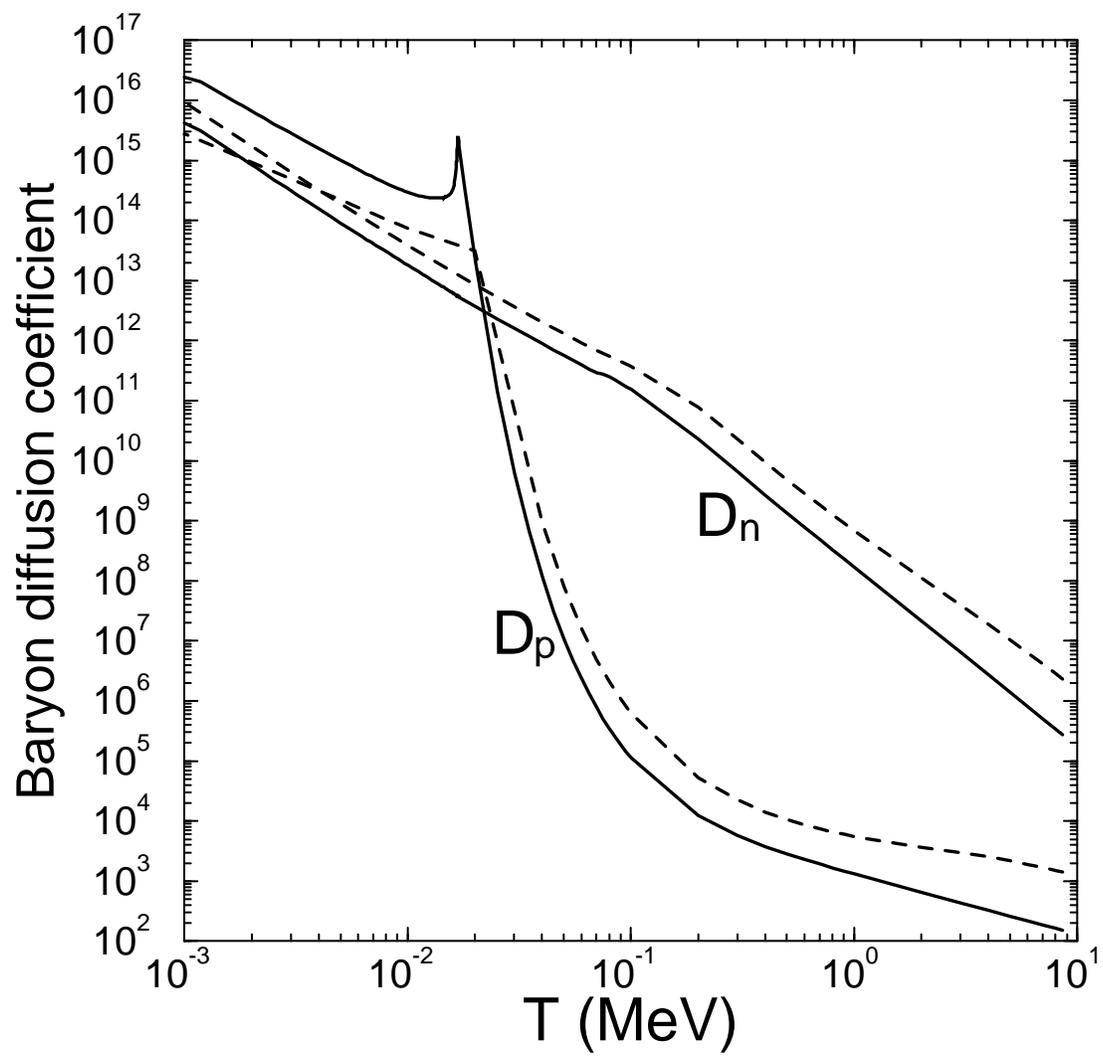
\item [{\rm Fig. 3}]
The thermal evolution of the neutron and proton diffusion coefficients
during the nucleosynthesis epoch for the fixed parameters,
$R = 10^6, \; f_{V}^{1/3} = 0.25, \; r = 100 \, m$,
and for the case $\omegabh2 = 0.1$.
The solid lines are results obtained using
the AHS diffusion coefficients, the dashed lines are obtained with 
finite temperature corrections included.
 
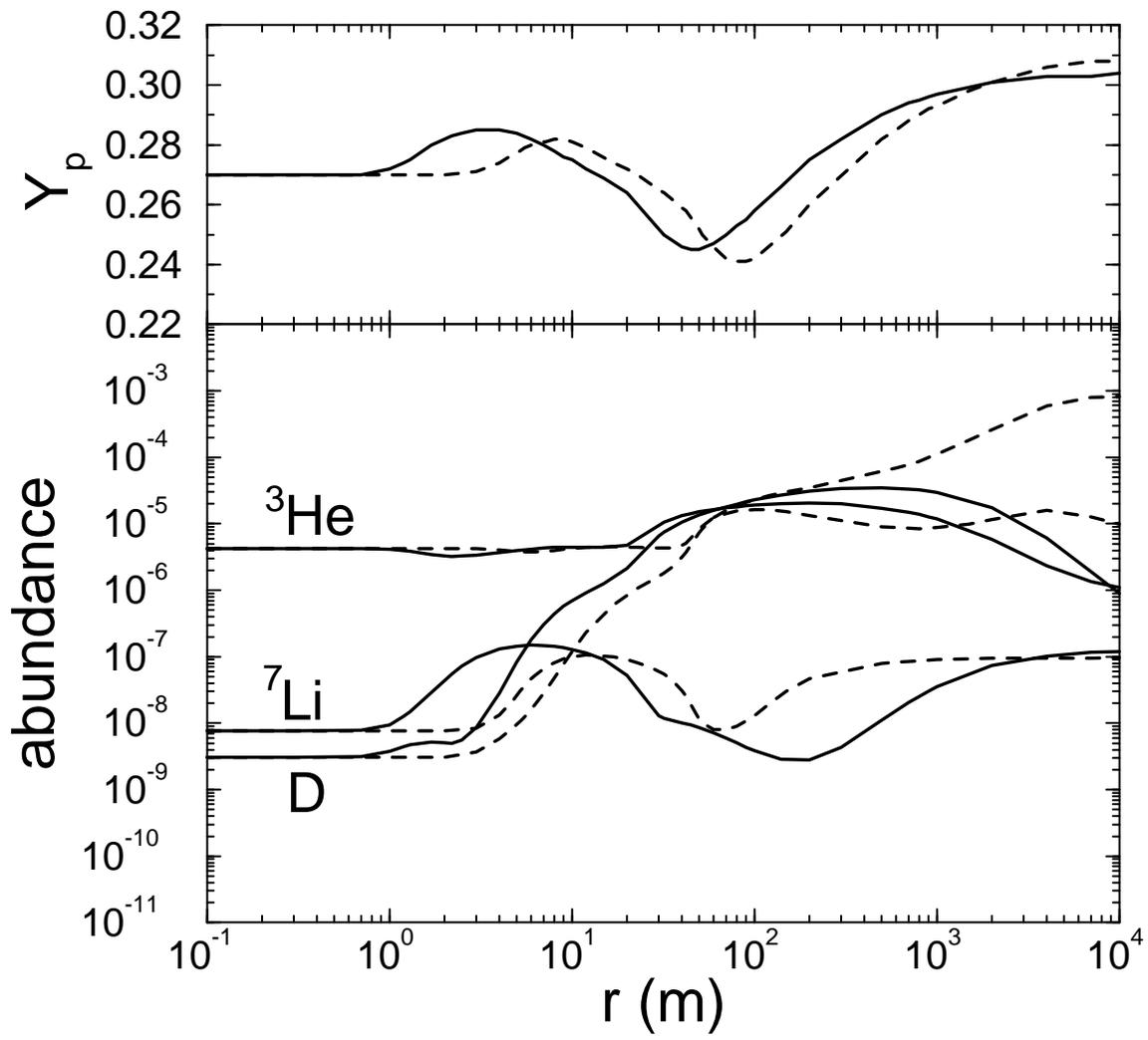
\item [{\rm Fig. 4}]
Light-element abundances as a function of the fluctuation distance scale 
for $\omegabh2 = 1.0$.
The initial baryon density ratio is $R = 10^{6}$ and the high density volume
fraction is $f_{V}^{1/3} = 0.25$. The solid lines are results obtained using
the AHS diffusion coefficients, the dashed lines are obtained with the
diffusion coefficients in which finite temperature corrections are included.
The fluctuation length scale $r$ is given in units of meters comoving at
$T = 100$ MeV. This calculation is based upon baryon density fluctuations
represented by condensed spheres.
 
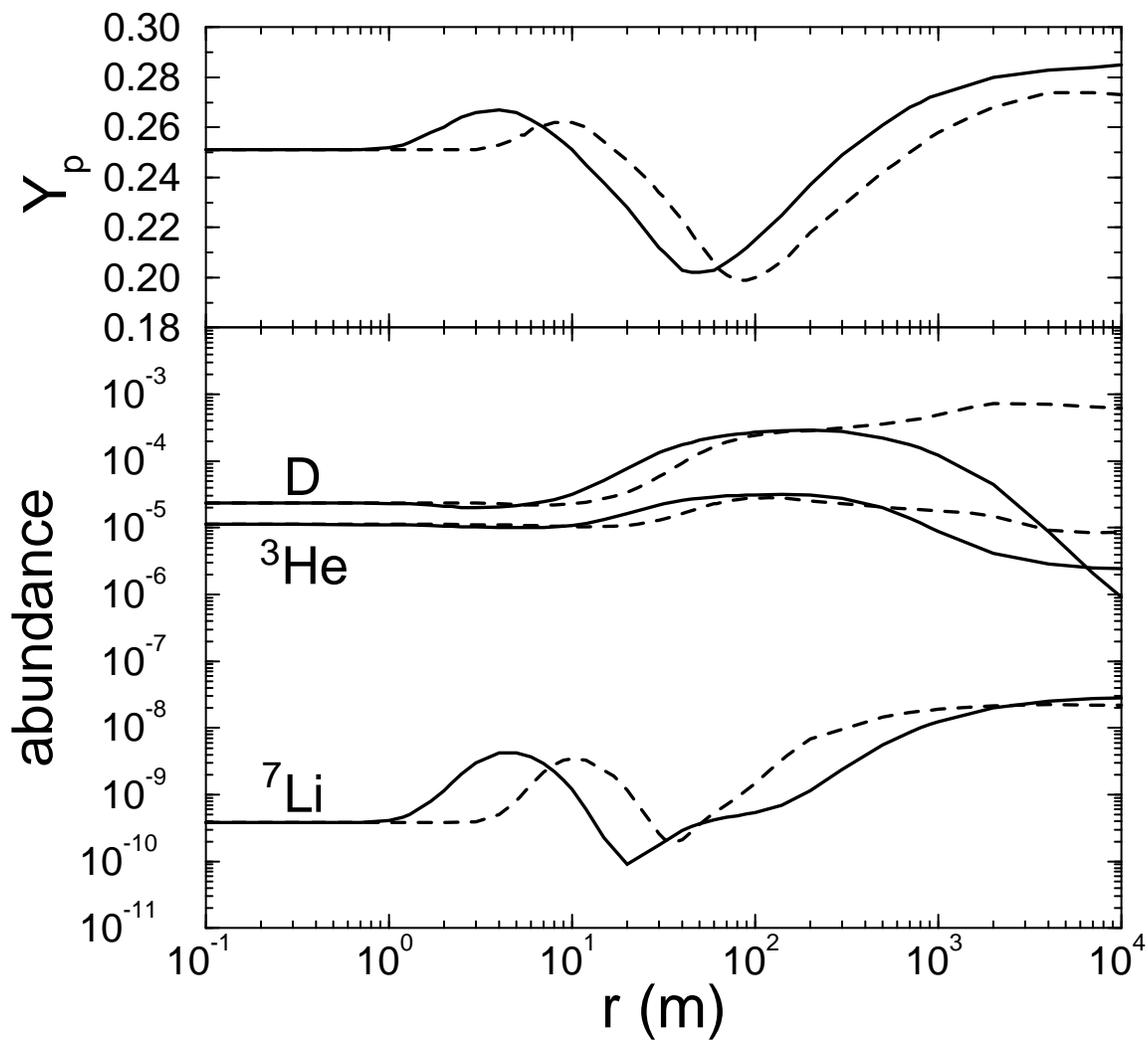
\item [{\rm Fig. 5}] Same as Fig. 4, but for $\omegabh2 = 0.1$
 
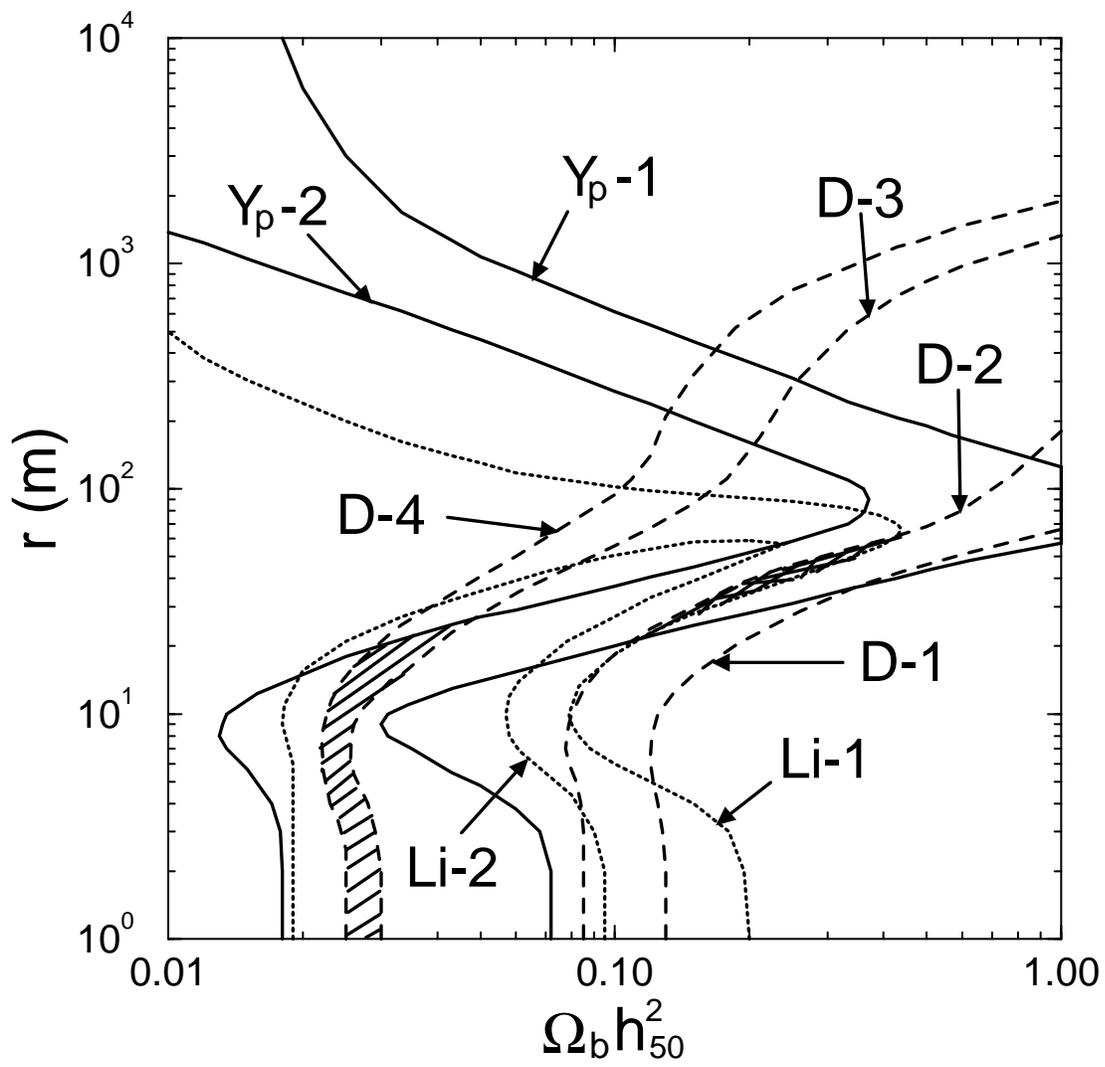
\item [{\rm Fig. 6}]
Allowed values for $\omegabh2$ and the fluctuation length scale $r$
based upon the various light-element abundance constraints as indicated.
The fluctuation length scale $r$ is given in units of meters comoving at
$T = 100$ MeV. This calculation is based upon baryon density fluctuations
represented by condensed spheres.
The adopted primordial light-element abundance constraints are
$Y_p$-1) $Y_p \leq 0.247$, $Y_p$-2) $Y_p \geq 0.226$;
D-1) $D/H \geq 1.5 \times 10^{-5}$, 
D-2) $D/H \leq 3.3 \times 10^{-5}$,
D-3) $D/H \geq 1.5 \times 10^{-4}$, 
D-4) $D/H \leq 2.3 \times 10^{-4}$;
Li-1) $^{7}Li/H \leq 1.5 \times 10^{-9}$,
Li-2) $^{7}Li/H \leq 3.5 \times 10^{-10}$.
The single hatched region corresponds to the region allowed by the larger $D$ 
constraints(D-3,D-4). The cross hatched region denotes the allowed region by 
smaller value of $D$ constraints(D-1,D-2) plus the larger $Li/H$ upper limit (Li-1).
  
\end{itemize}
 
\end{document}